\documentclass[11pt]{revtex4}
\usepackage{amssymb,epsf}
\usepackage{latexsym}

\begin{document}

\title{Wormhole solutions in the presence of nonlinear Maxwell field}
\author{S. H. Hendi$^{1,2}$\footnote{email address: hendi@shirazu.ac.ir}}
\affiliation{$^1$ Physics Department and Biruni Observatory, College of Sciences, Shiraz
University, Shiraz 71454, Iran\\
$^2$ Research Institute for Astronomy and Astrophysics of Maragha (RIAAM),
P.O. Box 55134-441, Maragha, Iran}

\begin{abstract}
Generalizing of the Maxwell field to nonlinear electrodynamics theories, we
look for the magnetic solutions. In initial approximation these models give
the usual linear electrodynamics. We consider a suitable metric and
investigate the properties of the solutions. Also, we use the cut-and-paste
method to construct wormhole structure. We generalize the static solutions
to rotating spacetime and obtain conserved quantities.
\end{abstract}

\maketitle

\section{Introduction}

A wormhole can be defined as a tunnel which can joint two
universes \cite{MorTho}. Since General Relativity does not
preclude the existence of (traversable) wormholes, a large number
of papers have been written which clarify, support, or contradict
much of the research about wormholes. Morris and Thorne
\cite{MorTho} have shown that in order to construct a traversable
wormhole, one needs to have extraordinary material, denoted as
exotic matter. Exotic matter can guarantee the flare-out condition
of the wormhole at its throat.

Unlike the classical form of matter \cite{HawkingLSS}, it is
believed that the exotic matter violates the well-known energy
conditions such as the null energy conditions (NEC), weak energy
conditions (WEC), strong energy conditions (SEC) and dominant
energy conditions (DEC). One of the open questions about the
exotic matter is that if it can be formed in macroscopic
quantities or not. We should note that these energy conditions are
violated by certain states of quantum fields, amongst which we may
refer to the Casimir energy, Hawking evaporation, and vacuum
polarization \cite{NegativeEnergy}. Furthermore, it has been shown
that one of the effective causes of the (late time) cosmic
acceleration is an exotic fluid \cite{Exotic1}. Hence, it will be
motivated to study wormhole solutions, at least geometrically.

Traversable wormholes in the Dvali-Gabadadze-Porrati theory with
cylindrical symmetry has been studied in Ref. \cite{Richarte}.
Higher dimensional Lorentzian wormholes have been analyzed by
several authors \cite{Kar}. Moreover, wormhole solutions of higher
derivative gravity with linear and nonlinear Maxwell fields have
been considered in \cite{WormHDG}. For other kinds of wormhole
solutions, we refer the reader to Refs. \cite{CUTpaste,WormHDG}
and references therein.

Many authors have extensively considered the nonlinear
electrodynamics and used their results to explain some physical
phenomena \cite{Cuesta,Marklund,BIpapers,PMIpapers,HendiJHEP}. A
charged system whose performance cannot be described by the linear
equations may be characterized with nonlinear electrodynamics.
From mathematical point of view, since Maxwell equations
originated from the empirical nature, we can consider a general
nonlinear theory of electrodynamics and state that the Maxwell
fields, are only approximations of nonlinear electrodynamics,
which the approximation breaks down for the small distances. From
physical viewpoint, generalizations of Maxwell theory to nonlinear
electrodynamics were introduced to eliminate infinite quantities
in theoretical analysis of the electrodynamics \cite{BIpapers}. In
addition, one may find some various limitations of the linear
electrodynamics in Ref. \cite{Delphenich}.

Recently, we take into account new classes of nonlinear electrodynamics,
such as Born-Infeld (BI) like \cite{HendiJHEP} and power-Maxwell invariant
(PMI) \cite{PMIpapers} nonlinear electrodynamics, in order to obtain new
analytical solutions in Einstein and higher derivative gravity.

Motivated above, in this paper we look for the analytical magnetic
horizonless solutions of Einstein gravity with nonlinear Maxwell source.
Properties of the solutions will be investigated.

\section{Field Equations and wormhole solutions:}

The field equations of Einstein gravity with an arbitrary $U(1)$ gauge field
as a source, may be written as%
\begin{equation}
G_{\mu \nu }+\Lambda g_{\mu \nu }=\frac{1}{2}g_{\mu \nu
}L(\mathcal{F})-2L_{\mathcal{F}}F_{\mu \lambda }F_{\nu
}^{\;\lambda },  \label{Geq}
\end{equation}
\begin{equation}
\partial _{\mu }\left( \sqrt{-g}L_{\mathcal{F}}F^{\mu \nu }\right) =0,
\label{Maxeq}
\end{equation}
where $G_{\mu \nu }$ is the Einstein tensor, $\Lambda =-3/2l^{2}$ denotes
the four dimensional negative cosmological constant, $L(\mathcal{F})$ is an
arbitrary function of the closed $2$-form Maxwell invariant $\mathcal{F}%
=F_{\mu \nu }F^{\mu \nu } $ and $L_{\mathcal{F}}=\frac{dL(\mathcal{F})}{d%
\mathcal{F}}$.

In addition to PMI and BI theories, in this paper, we take into account the
recently proposed BI-like models \cite{HendiJHEP}, which we called them as
Exponential form of nonlinear electrodynamics theory (ENE) and Logarithmic
form of nonlinear electrodynamics theory (LNE), whose Lagrangians are
\begin{equation}
L(\mathcal{F})=\left\{
\begin{array}{ll}
\left( -\mathcal{F}\right) ^{s}, & \;~{PMI} \\
4\beta ^{2}\left( 1-\sqrt{1+\frac{\mathcal{F}}{2\beta ^{2}}}\right), & \;~{BI%
} \\
\beta ^{2}\left( \exp (-\frac{\mathcal{F}}{\beta ^{2}} )-1\right) , & \;~{ENE%
} \\
-8\beta ^{2}\ln \left( 1+\frac{\mathcal{F}}{8\beta ^{2}} \right) , & \;~{LNE}%
\end{array}
\right. ,  \label{Lnon}
\end{equation}
where $s$ and $\beta $ are two nonlinearity parameters. Expanding the
mentioned Lagrangians near the linear Maxwell case ($s\longrightarrow 1$ and
$\beta \longrightarrow \infty $), one can obtain%
\begin{equation}
L(\mathcal{F})\longrightarrow L_{Max}+\left\{
\begin{array}{ll}
-\mathcal{F}\ln \left( -\mathcal{F}\right) (s-1)+O(s-1)^{2}, & \;~{PMI} \\
+\frac{\chi \mathcal{F}^{2}}{16\beta ^{2}}+O\left( \frac{\mathcal{F}^{3}}{%
\beta ^{4}}\right), & \;~{others}%
\end{array}
\right. ,  \label{Lbast}
\end{equation}
where Maxwell Lagrangian $L_{Max}=-\mathcal{F}$ and $\chi =1$, $2$ and $8$
for LNE, BI and ENE branches, respectively.

Investigation of the effects of the higher derivative corrections to the
gauge field seems to be an interesting phenomenon. These nonlinear
electrodynamics sources have different effects on the physical properties of
the solutions. For example in black hole framework, these nonlinearities may
change the electric potential, temperature, horizon geometry, energy density
distribution and also asymptotic behavior of the solutions. In what follows,
we study the effects of nonlinearity on the magnetic solutions.

\begin{figure}[tbp]
\epsfxsize=7cm \centerline{\epsffile{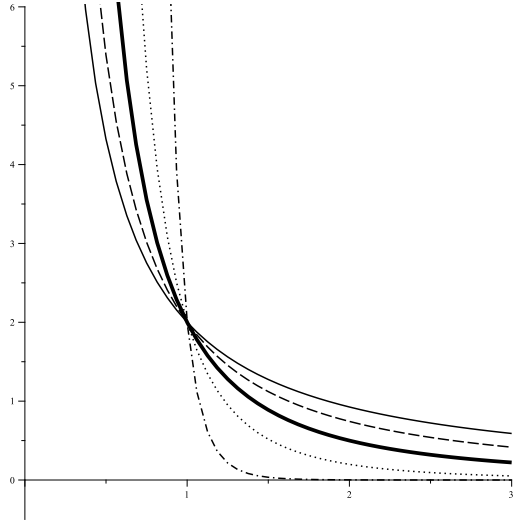}} \caption{$E(r)$
versus $r$ for $q=1$, $\Upsilon=1$, $l=1$ and $s=1.4$ (solid
line), $s=1.2$ (dashed line), $s=1$ "Maxwell field" (bold line),
$s=0.8$ (dotted line) and $s=0.6$ (dash-dotted line)} \label{Epmi}
\end{figure}

\begin{figure}[tbp]
\epsfxsize=7cm \centerline{\epsffile{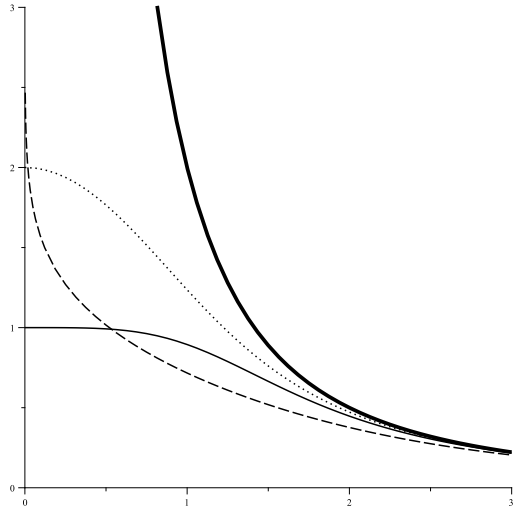}}
\caption{$E(r)$ versus $r$ for $q=1$, $\Upsilon=1$, $l=1$ and $\protect\beta%
=1$. BI (solid line), ENE (dashed line), LNE (dotted line) and Maxwell field
(bold line)}
\label{Eothers}
\end{figure}

Motivated by the fact that we are looking for the horizonless magnetic
solution (instead of electric one), one can start with the following $4$%
-dimensional spacetime
\begin{equation}
ds^{2}=-\frac{r^{2}}{l^{2}}dt^{2}+\frac{dr^{2}}{f(r)}+\Upsilon
^{2}l^{2}f(r)d\theta ^{2}+r^{2}d\phi ^{2},  \label{StaticMetric}
\end{equation}%
where $\Upsilon $ is a constant and will be fixed later. We should note
that, because of the periodic nature of $\theta $, one can obtain the
presented metric (\ref{StaticMetric}) with \emph{local} transformations $%
t\rightarrow il\Upsilon \theta $ and $\theta \rightarrow it/l$ in the
horizon-flat Schwarzschild metric, $ds^{2}=-f(r)dt^{2}+\frac{dr^{2}}{f(r)}%
+r^{2}\left( d\theta ^{2}+d\phi ^{2}\right) $. In other words, metric (\ref%
{StaticMetric}) may be \emph{locally} mapped to Schwarzschild spacetime, but
not \emph{globally}. Considering the mentioned local transformation, one can
find that the nonzero component of the gauge potential is $A_{\theta }$%
\begin{equation}
A_{\mu }=h(r)\delta _{\mu }^{\theta },  \label{PotStatic}
\end{equation}%
where $h(r)$ is an arbitrary function of $r$. Using Eq. (\ref{Maxeq}) with
the metric (\ref{StaticMetric}), we find $h(r)=\int E(r)dr$ in which%
\begin{equation}
E(r)=\left\{
\begin{array}{ll}
\frac{2ql^{2}\Upsilon ^{2}}{r^{2/(2s-1)}}, & \;~PMI \\
\frac{2ql^{2}\Upsilon ^{2}}{r^{2}\sqrt{1+\frac{4q^{2}l^{4}\Upsilon ^{2}}{%
\beta ^{2}r^{4}}}}, & \;~BI \\
\frac{2ql^{2}\Upsilon ^{2}}{r^{2}}\exp \left( -\frac{L_{W}}{2}\right) , & \;~%
{ENE} \\
\frac{\beta ^{2}r^{2}\left( \Gamma -1\right) }{ql^{2}}, & \;~{LNE}%
\end{array}%
\right. ,  \label{hp}
\end{equation}%
where $L_{W}=LambertW\left( X\right) ,$ $X=\frac{16l^{4}q^{2}\Upsilon ^{2}}{%
\beta ^{2}r^{4}}$ and $\Gamma =\sqrt{1+\frac{X}{4}}$ and therefore the
nonzero component of electromagnetic field tensor is%
\begin{equation}
F_{r\theta }=E(r).  \label{Frtheta}
\end{equation}%
We should note that the physical gauge potential should vanish for large
values of $r$. This condition is satisfied for $1/2<s<3/2$ and arbitrary $%
\beta $ (the mentioned constrain for $s$ is used throughout the rest of the
paper). Now, one can expand Eq. (\ref{hp}) to obtain the leading
nonlinearity correction of Maxwell field
\begin{equation}
\left. E(r)\right\vert _{\;near\;the\;linear\;case}=\frac{2ql^{2}\Upsilon
^{2}}{r^{2}}+\left\{
\begin{array}{ll}
\frac{8ql^{2}\Upsilon ^{2}\ln (r)}{r^{2}}(s-1)+O(s-1)^{2}, & \;~PMI \\
-\frac{2\chi q^{3}l^{6}\Upsilon ^{4}}{\beta ^{2}r^{6}}+O(\frac{1}{\beta ^{4}}%
), & \;others%
\end{array}%
\right. .  \label{Ebast}
\end{equation}%
In order to examine the effect of nonlinearity on the
electromagnetic field, we plot Figs. \ref{Epmi} and \ref{Eothers}.
Figure \ref{Epmi} shows that when we reduce the nonlinearity $s$,
the electromagnetic field of the PMI branch diverges for
$r\longrightarrow 0$ more rapidly and for large distances it goes
to zero more quickly. Figure \ref{Eothers} shows that for all
BI-like branches, the electromagnetic field (the same behavior as
in Maxwell case) vanishes for large $r$. Near the origin the
electromagnetic field of BI and LNE branches have finite values,
but for ENE branch, it diverges. Comparing this divergency with
Maxwell one, we find that ENE divergency is more slowly.


Taking into account the electromagnetic field tensor, we are in a position
to find the function $f(r)$. In order to obtain it, one may use any
components of Eq. (\ref{Geq}). We simplify the components of Eq. (\ref{Geq})
and find that the nonzero independent components of Eq. (\ref{Geq}) are
\begin{eqnarray}
&&f^{\prime \prime }(r)+\frac{2f^{\prime }(r)}{r}+2\Lambda +\Delta _{1}(r)
=0,  \label{FE1} \\
&&f^{\prime }(r)+\frac{f(r)}{r}+\Lambda r+\Delta _{2}(r) =0,  \label{FE2}
\end{eqnarray}
with%
\begin{eqnarray}
\Delta _{1}(r) &=&\left\{
\begin{array}{ll}
-\left( \frac{2h^{\prime 2}(r)}{\Upsilon ^{2}}\right) ^{s}, & \;~PMI \\
4\beta ^{2}\left[ \sqrt{1-\frac{h^{\prime 2}(r)}{\beta ^{2}\Upsilon ^{2}}}-1%
\right], & \;~BI \\
\beta ^{2}\left[ 1-\exp \left( \frac{2h^{\prime 2}(r)}{\beta ^{2}\Upsilon
^{2}}\right) \right] , & \;~{ENE} \\
8\beta ^{2}\ln \left[ 1-\left( \frac{h^{\prime }(r)}{2\beta \Upsilon }%
\right) ^{2}\right] , & \;~{LNE}%
\end{array}
\right. ,  \label{delta1} \\
\Delta _{2}(r) &=&\left\{
\begin{array}{ll}
\frac{r}{2}(2s-1)\left( \frac{2h^{\prime 2}(r)}{\Upsilon ^{2}}\right) ^{s},
& \;~PMI \\
2r\beta ^{2}\left[ \left( 1-\frac{h^{\prime 2}(r)}{\beta ^{2}\Upsilon ^{2}}%
\right) ^{-1/2}-1\right], & \;~BI \\
\frac{r\beta ^{2}}{2}\left\{ 1-\left[ 1-\left( \frac{2h^{\prime }(r)}{\beta
\Upsilon }\right) ^{2}\right] \exp \left( \frac{ 2h^{\prime 2}(r)}{\beta
^{2}\Upsilon ^{2}}\right) \right\}, & \;~{ENE} \\
4r\beta ^{2}\left\{ \ln \left[ 1-\left( \frac{h^{\prime }(r)}{2\beta
\Upsilon }\right) ^{2}\right] -\frac{2}{1-\left( \frac{2\beta \Upsilon }{%
h^{\prime }(r)}\right) ^{2}}\right\}, & \;~{LNE}%
\end{array}
\right. ,  \label{delta2}
\end{eqnarray}
where prime and double prime are first and second derivatives with respect
to $r$ , respectively. After some cumbersome manipulation, the solutions of
Eqs. (\ref{FE1}) and (\ref{FE2}) can be written as
\begin{equation}
f(r)=\frac{-2m}{r}-\frac{\Lambda r^{2}}{3}+\left\{
\begin{array}{ll}
-\frac{r^{2}(2s-1)^{2}}{2(2s-3)}\left( \frac{8q^{2}l^{4}\Upsilon ^{2}}{%
r^{4/(2s-1)}}\right) ^{s}, & \;~PMI \\
\frac{2\beta ^{2}r^{2}}{3}-\frac{2\beta ^{2}}{r}\int \sqrt{1+\frac{%
4q^{2}l^{4}\Upsilon ^{2}}{\beta ^{2}r^{4}}}r^{2}dr, & \;~BI \\
-\frac{\beta ^{2}r^{2}}{6}+\frac{2\beta ql^{2}\Upsilon }{r}\int \left( \frac{%
1}{\sqrt{L_{W}}}-\sqrt{L_{W}}\right) dr, & \;~{ENE} \\
+\frac{8\beta ^{2}r^{2}}{3}-\frac{4\beta ^{2}}{r}\int r^{2}\ln \left( \frac{%
\beta ^{2}r^{4}\left( \Gamma -1\right) }{2q^{2}\Upsilon ^{2}l^{4}}\right) dr-%
\frac{16l^{4}q^{2}\Upsilon ^{2}}{r}\int \frac{dr}{r^{2}\left( \Gamma
-1\right) }, & \;~{LNE}%
\end{array}
\right. ,  \label{f(r)}
\end{equation}
where $m$ is the integration constant which is related to the mass
parameter. In order to investigate the effect of nonlinearity on the metric
function, simplistically, we expand $f(r)$\ for $s\longrightarrow 1$ for PMI
and $\beta \longrightarrow \infty $ for other branches. After some
manipulation, we obtain
\begin{equation}
f(r)=f_{Max}(r)+\left\{
\begin{array}{ll}
\frac{4q^{2}l^{4}\Upsilon ^{2}\left[ 6+\ln \left( 8q^{2}\Upsilon
^{2}l^{4}r^{4}\right) \right] }{r^{2}}(s-1)+O(s-1)^{2}, & \;~PMI \\
-\frac{2\chi q^{4}l^{8}\Upsilon ^{4}}{5\beta ^{2}r^{6}}+O(\frac{1}{\beta ^{4}%
}), & \;others%
\end{array}
\right. ,  \label{Bast}
\end{equation}
where $f_{Max}(r)$ is the magnetic solution of Einstein-Maxwell gravity
\begin{equation}
f_{Max}(r)=\frac{-2m}{r}-\frac{\Lambda r^{2}}{3}+\frac{4q^{2}l^{4}\Upsilon
^{2}}{r^{2}},  \label{f(r)Maxwell}
\end{equation}
and the second term on the right hand side of Eq. (\ref{Bast}) is the
leading nonlinearity correction to the Einstein-Maxwell gravity solution.

\subsection{Properties of the solutions}

At first step, we should note that the presented solutions are
asymptotically anti-de Sitter (adS) and they reduce to asymptotically adS
Einstein-Maxwell solutions for $s\longrightarrow 1$ (PMI branch) or $\beta
\longrightarrow \infty $ (other branches).

The second step should be devoted to singularities and hence we should
calculate the curvature invariants. One can show that for the metric (\ref%
{StaticMetric}), the Kretschmann and Ricci scalars are
\begin{eqnarray}
R_{\mu \nu \rho \sigma }R^{\mu \nu \rho \sigma } &=&f^{\prime \prime 2}(r)+%
\frac{4f^{\prime 2}(r)}{r^{2}}+\frac{4f^{2}(r)}{r^{4}},  \label{Kretschmann}
\\
R &=&-f^{\prime \prime }(r)-\frac{4f^{\prime }(r)}{r}-\frac{2f(r)}{r^{2}}.
\label{Ricci}
\end{eqnarray}
Inserting Eq. (\ref{f(r)}) into the Eqs. (\ref{Kretschmann}) and (\ref{Ricci}%
), and using numerical calculations, one can show that the Ricci and
Kretschmann scalars diverge at $r=0$, are finite for $r>0$\ and for $%
r\rightarrow \infty $ one obtains
\begin{eqnarray}
\left. R_{\mu \nu \rho \sigma }R^{\mu \nu \rho \sigma }\right\vert _{
Large\; r} &=&\frac{8\Lambda ^{2}}{3}+\left\{
\begin{array}{ll}
O\left( \frac{1}{r^{\xi }}\right) & \;~PMI \\
O\left( \frac{1}{r^{6}}\right), & \;~others%
\end{array}
\right. ,  \label{RRinfinity} \\
\left. R\right\vert _{Large\;r} &=&4\Lambda +\left\{
\begin{array}{ll}
O\left( \frac{1}{r^{\xi }}\right) & \;~PMI \\
O\left( \frac{1}{r^{8}}\right), & \;~others%
\end{array}
\right. ,  \label{Rinfinity} \\
\xi &\in &\left( 3,\infty \right),  \nonumber
\end{eqnarray}
which confirms that the asymptotic behavior of the solutions is adS.
Considering the divergency of the Ricci and Kretschmann scalars at the
origin, one may think that there is a curvature singularity located at $r=0$%
. This singularity will be naked if the function $f(r)$\ has no real root
(singularity is not covered with a horizon) and we are not interested in it.
Therefore, we consider the case in which the function $f(r)$ has at least a
non-extreme positive real root. It is notable that the function $f(r)$\ is
negative for $r=r_{+}-\epsilon $ ($\epsilon $ is an infinitesimal number),
and positive for $r>r_{+}$\ where $r_{+}$\ is the largest positive real root
of $f(r)=0$. Negativity of the function $f(r)$ leads to an apparent change
of metric signature and it forces us to consider $r_{+}\leq r<$\ $\infty $.
We should state that although the metric function $f(r)$\ vanishes at $%
r=r_{+}$, but we have $f^{\prime }(r=r_{+})\neq 0$ ($f^{\prime }(r=r_{+})>0$%
). In addition, there is no curvature singularity in the range $r_{+}\leq
r<\infty $. Following the procedure of Ref. \cite{HendiCQG}, one may find
that there is a conic singularity at $r=r_{+}$.

Removing this conical singularity with $\Upsilon =1/[lf^{\prime }(r_{+})]$
\cite{HendiCQG}, we desire to interpret the obtained solutions as wormholes.
In order to construct wormholes from the gluing, one requires to use the
cut-and-paste prescription \cite{CUTpaste}. In this method, we take into
account two geodesically incomplete copies of the solutions (removing from
each copy the forbidden region $\Omega $) with two copies of the boundaries $%
\partial \Omega $, where
\begin{eqnarray}
\Omega &\equiv &\left\{ r\mid r<r_{+}\right\} ,  \label{forbidden} \\
\partial \Omega &\equiv &\left\{ r\mid r=r_{+}\right\} .  \label{Bforbidden}
\end{eqnarray}%
Now, we identify two copies of the mentioned boundaries to build a
geodesically complete manifold. This cut-and-paste method constructs a
wormhole with a throat at $r=r_{+}$. In order to confirm this claim, we
should check the so-called flare-out condition at the throat. To do this,
one can consider a $2$-dimensional submanifold of the metric (\ref%
{StaticMetric}), $ds_{2-\dim }^{2}$, (with constant $t$ and $\theta $) and
embed it in a $3$-dimensional Euclidean flat space in cylindrical
coordinates, $ds_{3-\dim }^{2}$, where
\begin{eqnarray}
ds_{2-\dim }^{2} &=&\frac{dr^{2}}{f(r)}+r^{2}d\phi ^{2},  \label{2dim} \\
ds_{3-\dim }^{2} &=&dr^{2}+r^{2}d\phi ^{2}+dz^{2}.  \label{3dim}
\end{eqnarray}
Regarding the surface $z=z(r)$, we obtain
\begin{eqnarray}
\left. \frac{dr}{dz}\right\vert _{r=r_{+}} &=&\left. \sqrt{\frac{f(r)}{1-f(r)%
}}\right\vert _{r=r_{+}}=0,  \label{drdz} \\
\left. \frac{d^{2}r}{dz^{2}}\right\vert _{r=r_{+}} &=&\left. \frac{f^{\prime
}}{2\left[ 1-f\right] ^{2}}\right\vert _{r=r_{+}}=\frac{1}{2}f^{\prime
}(r=r_{+})>0,  \label{d2rdz2}
\end{eqnarray}
which shows that the flare-out condition may be satisfied for the surface $%
z=z(r)$ and therefore $r=r_{+}$ is the radius of the wormhole throat.

\begin{figure}[tbp]
$%
\begin{array}{cc}
\epsfxsize=7cm \epsffile{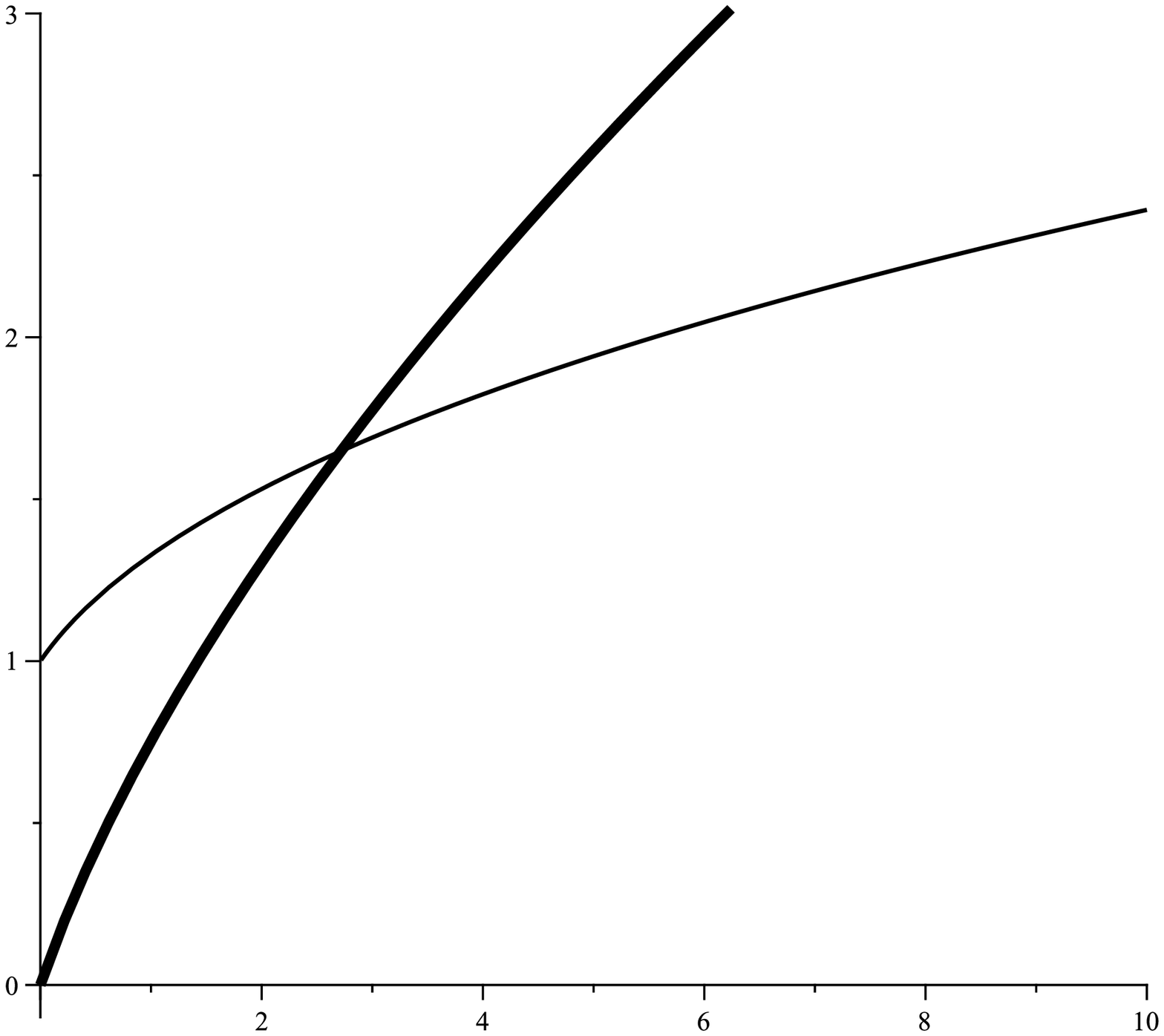} & \epsfxsize=7cm \epsffile{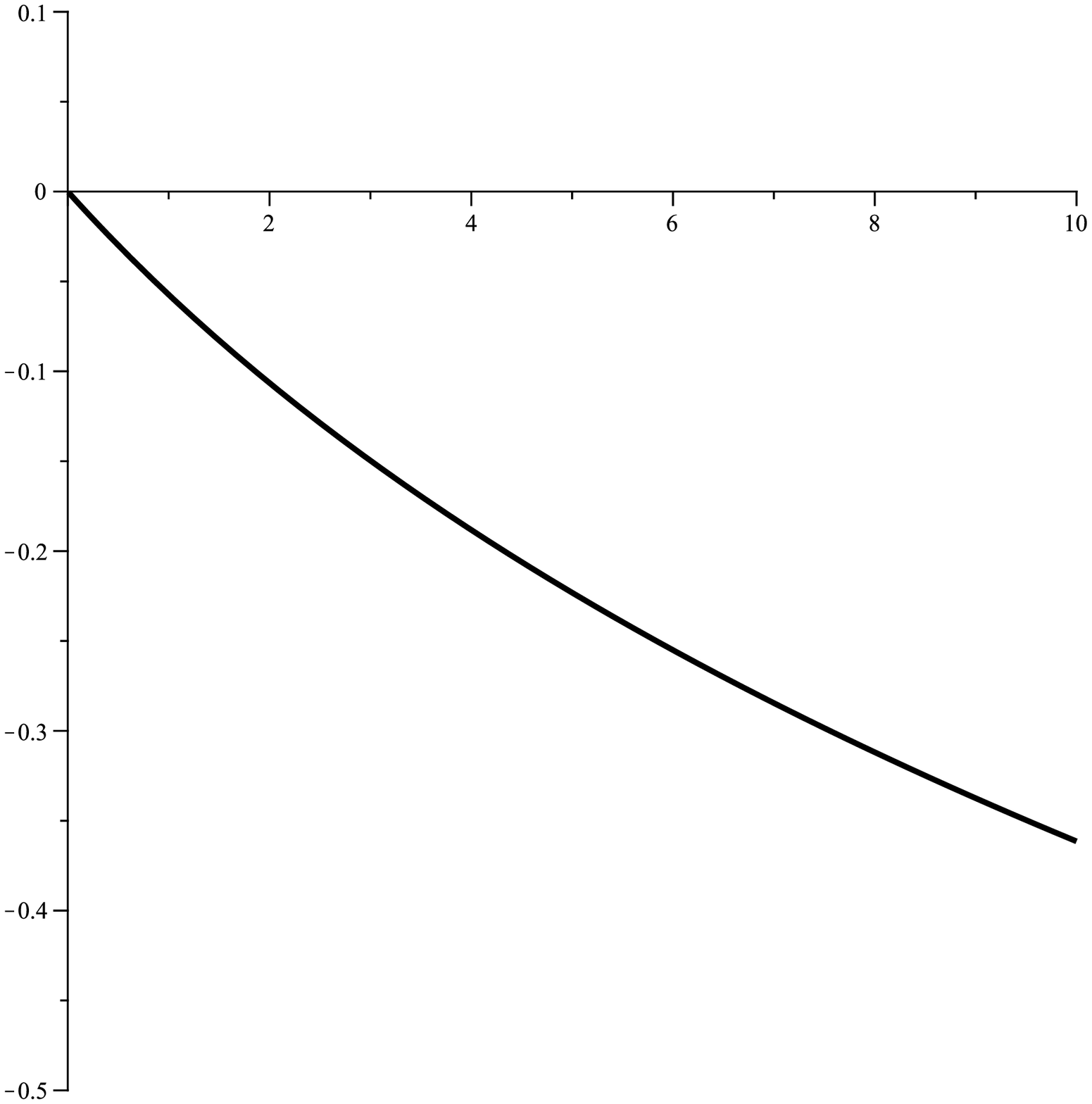}%
\end{array}
$%
\caption{\textbf{left figure:} $\protect\sqrt{\frac{L_{W}}{X}}$ (solid line) and $\protect\sqrt{%
XL_{W}}$ (bold line) versus $X$. \textbf{right figure:} $\ln
\left( \frac{8(\Gamma -1)}{X}\right) $ versus $X$.} \label{LWLN}
\end{figure}

Now, we should discuss the energy conditions for the wormhole solutions. We
should note that, traversable wormhole may exist with exotic matter which
violates the null energy condition \cite{MorTho}. We use the following
orthonormal contravariant (hatted) basis to simplify the mathematics and
physical interpretations
\begin{equation}
\mathbf{e}_{\widehat{t}}=\frac{l}{r}\frac{\partial }{\partial t},\;~ \;~
\mathbf{e}_{\widehat{r}}=f^{1/2}\frac{\partial }{\partial r},\;~\;~ \mathbf{e%
}_{\widehat{\theta }}=\frac{1}{\Upsilon lf^{1/2}}\frac{\partial }{\partial
\theta },\;~\;~\mathbf{e}_{\widehat{\phi }}=r^{-1}\frac{\partial }{\partial
\phi }.  \label{base}
\end{equation}
Using the mentioned basis, we can obtain
\begin{eqnarray}
T_{_{\widehat{t}\widehat{t}}} &=&-T_{_{\widehat{\phi }\widehat{\phi }
}}=\left\{
\begin{array}{ll}
-\frac{1}{2}\left( \frac{8\Upsilon ^{2}q^{2}l^{4}}{r^{4/(2s-1)}}\right) ^{s},
& \;~PMI \\
2\beta ^{2}\left( \Gamma ^{-1}-1\right), & \;~BI \\
\frac{\beta ^{2}}{2}\left( 1-\sqrt{\frac{X}{L_{W}}} \right), & \;~{ENE} \\
4\beta ^{2}\ln \left( \frac{8\left( \Gamma -1\right) }{X} \right), & \;~{LNE}%
\end{array}
\right. ,  \label{Tab1} \\
\;T_{_{\widehat{r}\widehat{r}}} &=&-T_{_{\widehat{\theta }\widehat{\theta }%
}}=\left\{
\begin{array}{ll}
-\frac{(2s-1)}{2}\left( \frac{8\Upsilon ^{2}q^{2}l^{4}}{r^{4/(2s-1)}}\right)
^{s}, & \;~PMI \\
2\beta ^{2}\left( 1-\Gamma \right), & \;~BI \\
\frac{\beta ^{2}}{2}\left( \sqrt{\frac{X}{L_{W}}}-\sqrt{ XL_{W}}-1\right), &
\;~{ENE} \\
\beta ^{2}\left[ 8-4\ln \left( \frac{8\left( \Gamma -1\right) }{X}\right) -%
\frac{X}{\left( \Gamma -1\right) }\right], & \;~{LNE}%
\end{array}
\right. ,  \label{Tab2}
\end{eqnarray}
and therefore
\begin{eqnarray}
T_{_{\widehat{t}\widehat{t}}}&<&0  \nonumber \\
T_{_{\widehat{t}\widehat{t}}}+T_{_{\widehat{r}\widehat{r}}}&<&0,  \nonumber
\end{eqnarray}
which shows that all the energy conditions are violated as well
(see Fig. \ref{LWLN} for more clarification).

\begin{figure}[tbp]
$%
\begin{array}{cc}
\epsfxsize=7cm \epsffile{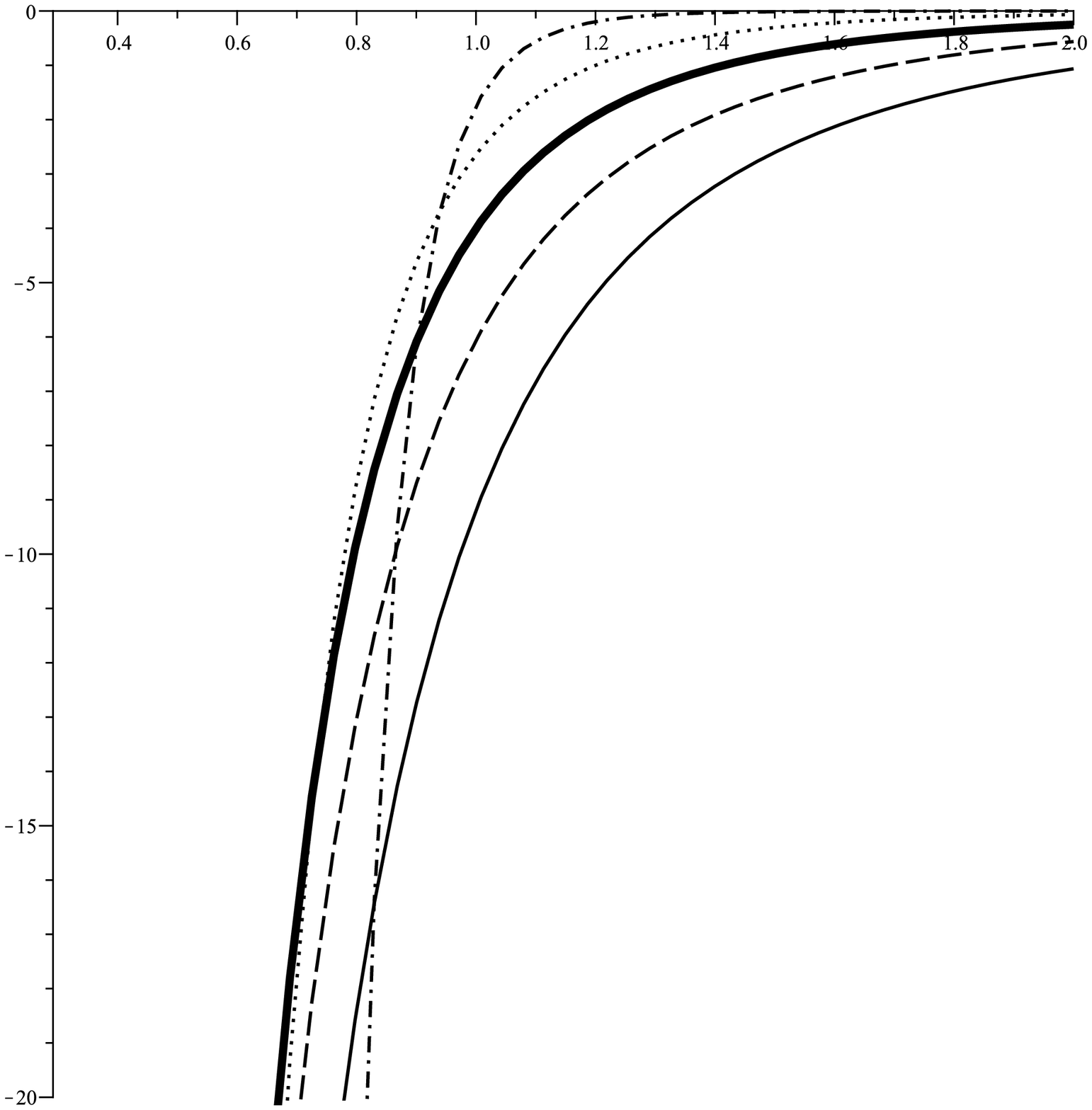} & \epsfxsize=7cm \epsffile{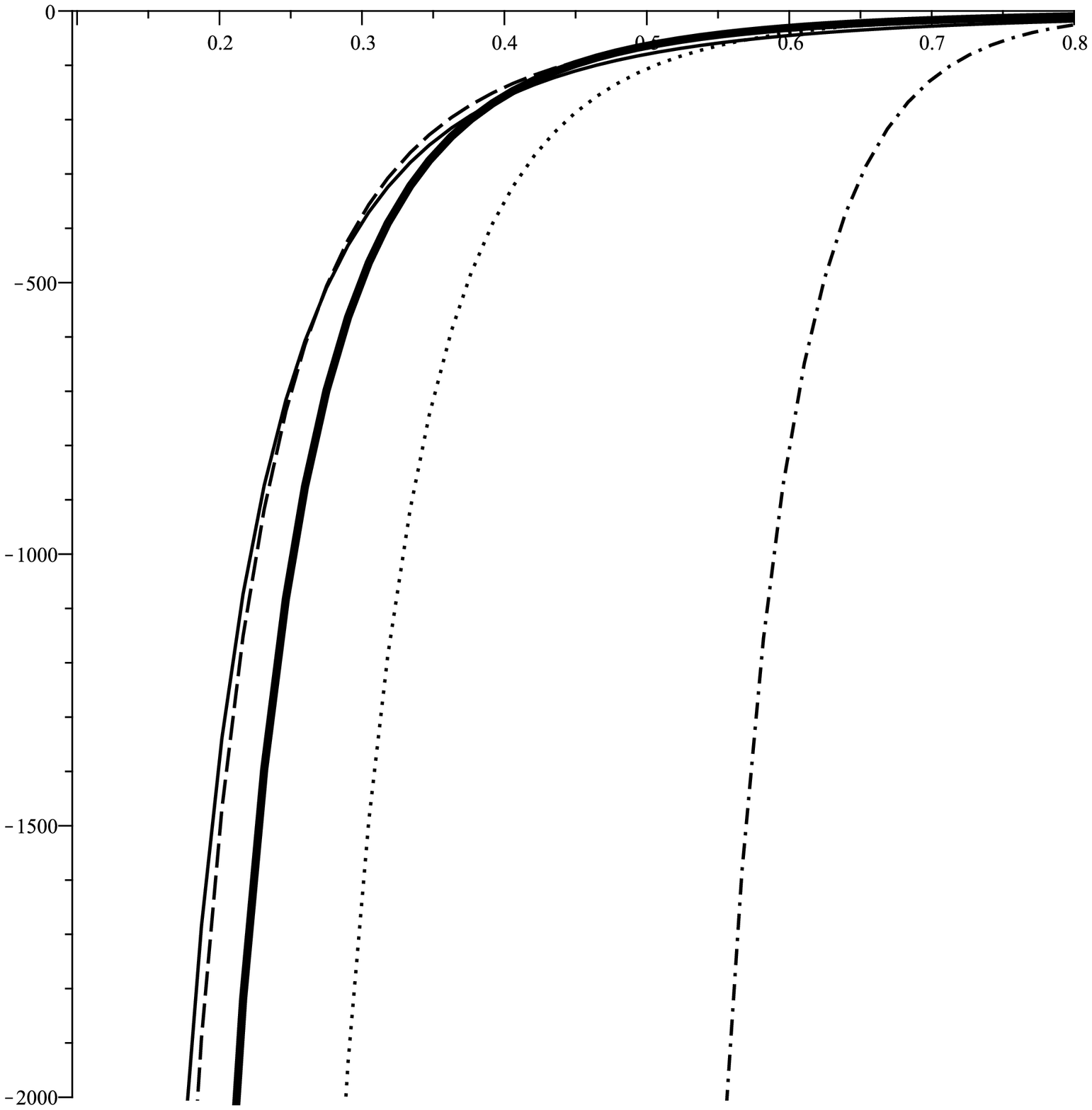}%
\end{array}
$%
\caption{$T_{_{\widehat{t}\widehat{t}}}$ versus $r$ for $q=1$, $\Upsilon=1$,
$l=1$ and $s=1.4$ (solid line), $s=1.2$ (dashed line), $s=1$ "Maxwell field"
(bold line), $s=0.8$ (dotted line) and $s=0.6$ (dash-dotted line) "\textbf{%
different scales}"}
\label{TttPMI}
\end{figure}

\begin{figure}[tbp]
\epsfxsize=7.5cm \centerline{\epsffile{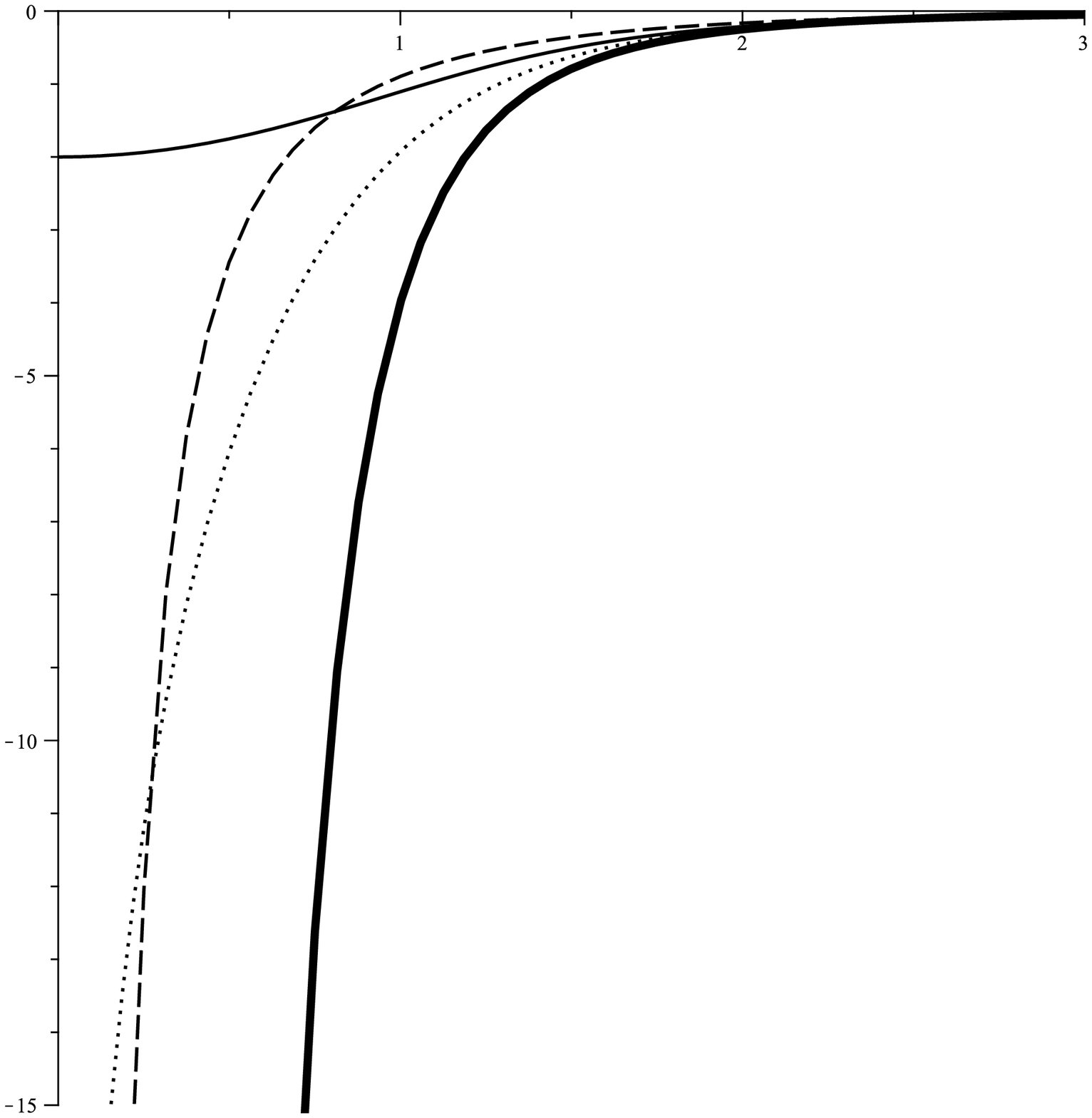}}
\caption{$T_{_{\widehat{t}\widehat{t}}}$ versus $r$ for $q=1$, $\Upsilon=1$,
$l=1$ and $\protect\beta=1$. BI (solid line), ENE (dashed line), LNE (dotted
line) and Maxwell field (bold line)}
\label{TttOTHERS}
\end{figure}

At the end of this section, we desire to study the effects of the
nonlinearity on energy density of the spacetime. At the start, we can expand
$T_{_{\widehat{t}\widehat{t}}}$ near the linear case to obtain
\begin{equation}
T_{_{\widehat{t}\widehat{t}}}=\left. T_{_{\widehat{t}\widehat{t}
}}\right\vert _{Maxwell}+\left\{
\begin{array}{ll}
-\frac{4q^{2}l^{4}\Upsilon ^{2}\ln \left( 8q^{2}\Upsilon
^{2}l^{4}r^{4}\right) }{r^{4}}(s-1)+O(s-1)^{2}, & \;~PMI \\
+\frac{6\chi q^{4}l^{8}\Upsilon ^{4}}{\beta ^{2}r^{8}}+O(\frac{1}{\beta ^{4}}%
), & \;others%
\end{array}
\right. ,  \label{Tbast}
\end{equation}
where $\left. T_{_{\widehat{t}\widehat{t}}}\right\vert _{Maxwell}=\frac{%
-4\Upsilon ^{2}q^{2}l^{4}}{r^{4}}$ and the second term on the right hand
side of Eq. (\ref{Tbast}) is the leading nonlinearity correction to the
energy density of the Einstein-Maxwell theory. In addition, we plot the
energy density $T_{_{\widehat{t}\widehat{t}}}$\ versus $r$ for different
values of nonlinearity parameter $s$ and also various branches of BI-like
fields. Figures \ref{TttPMI} and \ref{TttOTHERS}, show that for the
arbitrary choices of $r$ the energy density is negative. Furthermore, Fig. %
\ref{TttPMI} shows that the nonlinearity parameter, $s$, has effects on the
behavior of the energy density and when we reduce $s$, both divergency of
energy density near the origin and its vanishing for large values of
distance occur more rapidly. Moreover, considering Fig. \ref{TttOTHERS}, one
can find that $T_{_{\widehat{t}\widehat{t}}}$ has a finite value for an
arbitrary distance in BI branch and it diverges near the origin for other
branches. It is notable that, near the origin, the divergency of LNE branch
is stronger than ENE branch. Also, for BI-like branches, the nonlinearity
reduces the strength of energy density divergency.

\subsection{Rotating solutions \label{Rotworm}}

In this section, we want to add angular momentum to the static spacetime (%
\ref{StaticMetric}). To do this, one can use the following rotation boost in
the $t-\theta $ plane
\begin{equation}
t\mapsto \Xi t-a\theta ,\hspace{0.5cm}\theta \mapsto \Xi \theta -\frac{a}{%
l^{2}}t,  \label{Tr}
\end{equation}
where $\Xi =\sqrt{1+a^{2}/l^{2}}$ and $a$ is a rotation parameter. Taking
into account Eq. (\ref{Tr}) and applying it to static metric (\ref%
{StaticMetric}), one obtains
\begin{equation}
ds^{2}=-\frac{r^{2}}{l^{2}}\left( \Xi dt-ad\theta \right) ^{2}+\frac{dr^{2}}{%
f(r)}+\Upsilon ^{2}l^{2}f(r)\left( \frac{a}{l^{2}}dt-\Xi d\theta \right)
^{2}+r^{2}d\phi ^{2},  \label{RotatingMetric}
\end{equation}%
where $f(r)$ is the same as $f(r)$ given in Eq. (\ref{f(r)}). It is notable
that one can obtain the presented metric (\ref{RotatingMetric}) with \emph{%
local} transformations $t\rightarrow il\Upsilon \left( at/l^{2}-\Xi \theta
\right) $ and $\theta \rightarrow i\left( \Xi t-a\theta \right) /l$ in the
horizon-flat Schwarzschild metric, $ds^{2}=-f(r)dt^{2}+\frac{dr^{2}}{f(r)}%
+r^{2}d\theta ^{2}+r^{2}d\phi ^{2}$. Thus, the nonzero components of the
gauge potential are $A_{\theta }$ and $A_{t}$%
\begin{equation}
A_{\mu }=h(r)\left( \Xi \delta _{\mu }^{\theta }-\frac{a}{l^{2}}\delta _{\mu
}^{t}\right) ,  \label{PotRotating}
\end{equation}%
where $h(r)$ is the same as in the static case. Furthermore, the nonzero
components of electromagnetic field tensor are given by
\begin{equation}
F_{tr}=\frac{a}{\Xi l^{2}}F_{r\theta }=\frac{a}{l^{2}}E(r).  \label{Ftr}
\end{equation}%
As we mentioned before, the periodic nature of $\theta $ helps us to
conclude that the transformation (\ref{Tr}) is not a proper coordinate
transformation on the entire manifold and therefore the metrics (\ref%
{StaticMetric}) and (\ref{RotatingMetric}) are distinct \cite{Sta}. In
addition, it is desired to note that rotating solutions have no horizon and
curvature singularity. Moreover, it is worth noting that besides the
magnetic field along the $\theta $ coordinate, there is also a radial
electric field ($F_{tr}\neq 0$) and therefore, unlike the static case, the
rotating wormhole has a nonzero electric charge which is proportional to the
rotation parameter.

\subsection{Conserved Quantities \label{Conserve}}

Here we desire to calculate finite conserved quantities. In order to obtain
a finite value for these quantities, we can use the counterterm method
inspired by the concept of (AdS/CFT) correspondence \cite{Mal}. It has been
shown that for asymptotically AdS solutions the finite energy momentum
tensor is%
\begin{equation}
T^{ab}=\frac{1}{8\pi }\left( K^{ab}-K\gamma ^{ab}-\frac{2\gamma ^{ab}}{l}%
\right) ,  \label{Stress}
\end{equation}%
where $K$ is the trace of the extrinsic curvature $K^{ab}$ and $\gamma ^{ab}$
is the induced metric of the boundary. Taking into account the Killing
vector field $\mathcal{\xi }$, one may obtain the quasilocal conserved
quantities in the following form
\begin{equation}
\mathcal{Q}(\mathcal{\xi )}=\int_{\mathcal{B}}d^{2}\varphi \sqrt{\sigma }%
T_{ab}n^{a}\mathcal{\xi }^{b},  \label{charge}
\end{equation}%
where $\sigma $ is the determinant of the boundary metric in ADM
(Arnowitt-Deser-Misner) form $\sigma _{ij}$, and $n^{a}$ is the timelike
unit vector normal to the boundary $\mathcal{B}$. Considering two Killing
vectors $\xi =\partial /\partial t$ and $\zeta =\partial /\partial \theta $,
we can find their associated conserved charges which are mass and angular
momentum
\begin{equation}
{M}=4\pi ^{2}\left[ 3\left( \Xi ^{2}-1\right) +1\right] \Upsilon m,
\label{M}
\end{equation}
\begin{equation}
{J}=12\pi ^{2}\Upsilon m\Xi a,  \label{J}
\end{equation}
where the former equation confirms that $a$ is the rotational parameter.

Finally, we are in a position to discuss the electric charge. In order to
compute it, we need a nonzero radial electric field $F_{tr}$ and therefore
one expects vanishing $F_{tr}$ (static case) leads to zero electric charge.
Taking into account the Gauss's law for the rotating solutions and computing
the flux of the electric field at infinity, one can find
\begin{equation}
Q=\left\{
\begin{array}{ll}
\frac{2^{3s+1}\pi ^{2}\Upsilon s}{4l}\left( \frac{(2s-1)q}{(3-2s)l}\right)
^{2s-1}a, & \;~PMI \\
\frac{4\pi ^{2}\Upsilon q}{l^{2}}a, & \;~others%
\end{array}
\right. ,  \label{chden}
\end{equation}
which confirms that the static wormholes do not have electric charge.

\section{ Closing Remarks}

In this paper, we took into account a class of magnetic Einsteinian
solutions in the presence of nonlinear source. The magnetic spacetime which
we used in this paper, may be obtained from the horizon-flat Schwarzschild
metric with \emph{local} transformations $t\rightarrow il\Upsilon \theta $
and $\theta \rightarrow it/l$. It is notable that because of the periodic
nature of $\theta$, the mentioned transformations cannot be global.

We considered four forms of nonlinear electrodynamics, namely PMI, BI, ENE
and LNE theories, whose asymptotic behavior leads to Maxwell theory. We
investigated the effect of nonlinearity parameter on the electromagnetic
field and found that for PMI branch, if one reduces the nonlinearity
parameter $s$, then the electromagnetic field diverges near the origin more
rapidly and for large distances it goes to zero more quickly. In addition,
we found that for all BI-like branches, the behavior of the electromagnetic
field is the same as Maxwell case for large values of distance, but near the
origin, the electromagnetic field of the BI and LNE branches is finite and
it diverges for the ENE branch. It is interesting to note that the
divergency of the ENE branch has less strength in comparison to the the
Maxwell field.

Then, we obtained the metric function for all branches and found that they
reduce to asymptotically adS Einstein-Maxwell solutions for $%
s\longrightarrow 1$ (PMI branch) or $\beta \longrightarrow \infty $ (other
branches). We also expanded the metric function near the linear Maxwell
field and calculated the curvature scalars for large $r$ to find that
obtained solutions are asymptotically anti-de Sitter (adS). Taking into
account the presented metric, one can find that the function $f(r)$ cannot
be negative since its negativity leads to an apparent change of metric
signature. This limitation forced us to consider $r_{+}\leq r<$\ $\infty $.
Using numerical calculations, one can find that there is no curvature
singularity in the range $r_{+}\leq r<\infty $, but one may find a conic
singularity at $r=r_{+}$.

After that, we removed the mentioned conic singularity and used the
cut-and-paste prescription to construct a wormhole from the gluing and then
we checked the so-called flare-out condition at the throat $r=r_{+}$. Since
it has been stated before, that traversable wormhole may exist with exotic
matter \cite{MorTho}, we investigated the energy conditions for the obtained
wormhole solutions and found that the energy conditions are violated.

We also studied the effects of nonlinearity parameter on the energy density.
We found that when we reduce $s$, both divergency of energy density near the
origin and its vanishing for large values of distance occur more rapidly.
Moreover, one can find that energy density has a finite value for an
arbitrary distance only in BI branch and it diverges near the origin for
other BI-like branches. It is notable that, near the origin, the divergency
of LNE branch is stronger than ENE branch.

We generalized the static solutions to rotating ones and obtained the
conserved quantities. We found that, unlike the static case, for the
spinning spacetime, the wormhole has a net electric charge density. We also
found that in spite of the fact that the mentioned nonlinear theories change
the properties of the solutions significantly, but they do not have any
effect on mass and angular momentum.

\begin{acknowledgements}
We are indebted to H. Mohammadpour for reading the manuscript. We
also wish to thank Shiraz University Research Council. This work
has been supported financially by Research Institute for Astronomy
\& Astrophysics of Maragha (RIAAM), Iran.
\end{acknowledgements}

\end{document}